\documentclass[journal=jpclcd,manuscript=letter]{achemso}

\usepackage{chemformula} 
\usepackage[separate-uncertainty=true,separate-uncertainty,multi-part-units=single]{siunitx} 
\usepackage{array}
\usepackage{svg}

\author{S. Raaijmakers}
\author{M. Pols}
\author{J. M. Vicent-Luna*}
\email{j.vicent.luna@tue.nl}
\author{S. Tao*}
\email{s.x.tao@tue.nl}
    \affiliation{Materials Simulation \& Modelling, Department of Applied Physics, Eindhoven University of Technology, 5600 MB, Eindhoven, The Netherlands}
    \alsoaffiliation{Center for Computational Energy Research, Department of Applied Physics, Eindhoven University of Technology, 5600 MB, Eindhoven, The Netherlands}

\title{A Reparameterized Density Functional Tight-Binding Method for Engineering phase-stable CsPbX\textsubscript{3} Perovskites}

\begin{document}


\makeatletter
\setlength\acs@tocentry@height{5.75cm}
\setlength\acs@tocentry@width{7.75cm}
\makeatother

\begin{tocentry}

\includegraphics{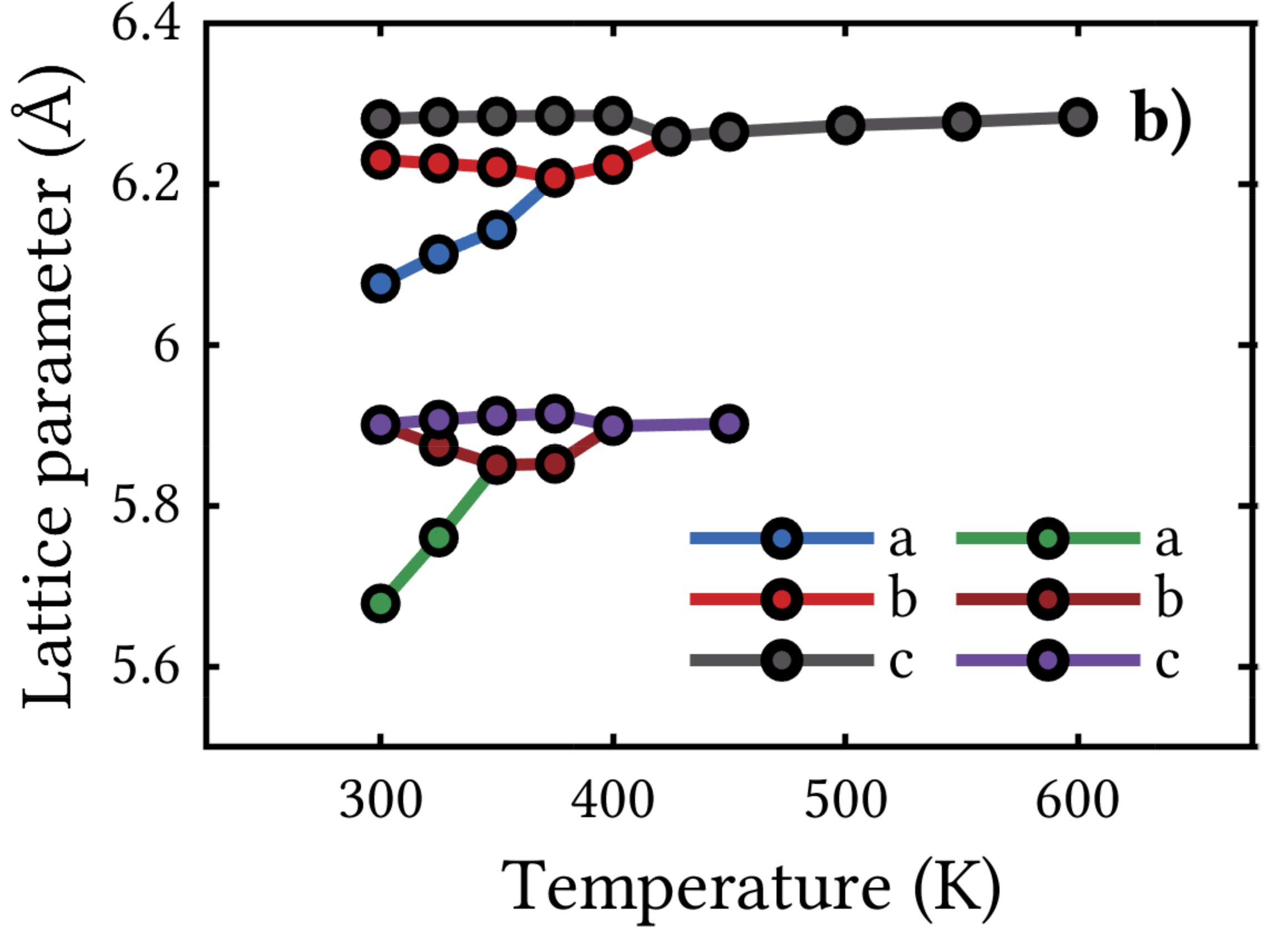}

\end{tocentry}

\begin{abstract}

Halide perovskites are a promising class of materials for optoelectronic applications, due to their excellent optoelectronic performance. However, they suffer several dynamical degradation problems, the characterization of which is challenging in experiments. Atomic scale simulations can provide valuable insights, however, the high computational cost of traditional quantum mechanical methods such as DFT makes it difficult to model dynamical processes in large perovskite systems. In this work, we present a re-parameterized GFN1-xTB method for the accurate description of structural and dynamical properties of CsPbBr\textsubscript{3}, CsPbI\textsubscript{3}, and CsPb(I\textsubscript{1-x}Br\textsubscript{x})\textsubscript{3}. Our molecular dynamics simulations show that the phase stability is strongly correlated to the displacement of ions in the perovskites. In the low temperature orthorhombic phase, the directional movement of the Cs cations decreases contact with the surrounding halides, initiating a transition to the non-perovskite phase. However, this loss of contact can be compensated by increased halide displacement, once enough thermal energy is available, resulting in a transition to the tetragonal or cubic phases. Furthermore, we find the mixing of halides increase halide displacement over a significant range of temperatures, resulting in lower phase transition temperatures and therefore improved phase stability.

\end{abstract}


\section{Manuscript}
Over the last decade, metal halide perovskites (MHPs) have emerged as a promising class of materials for use in solar cells and light-emitting diodes, thanks to their excellent optoelectronic properties \cite{Bi2016,Asghar2017,effrev,Jena2019,Lin2018}. Reported power conversion efficiencies have seen a rapid increase from merely 3.8\% (2009 \cite{Kojima2009}) to 25.5\% (2020 \cite{Yoo2021}), however, long-term stability issues remain a bottleneck for their use in commercial applications \cite{Petrus2017,Niu2015,Wang2019}. MHPs are three dimensional crystals with chemical formula AMX\textsubscript{3}, where A is an organic or inorganic monovalent cation (FA\textsuperscript{+}, MA\textsuperscript{+}, Cs\textsuperscript{+}), M is a metal divalent cation (Pb\textsuperscript{2+}, Sn\textsuperscript{2+}, Ge\textsuperscript{2+}), and X a halide anion (I\textsuperscript{-}, Br\textsuperscript{-}, Cl\textsuperscript{-}). The metal cations and halide anions form a framework of corner sharing MX\textsubscript{6} octahedra, with the remaining cuboctahedral cavities occupied by the monovalent A cations. The resulting material is physically soft and dynamic in nature \cite{Wang2019,Lee2021,Lai2018,Chu2020}, significantly affecting its optoelectronic performance and long-term stability.

One of the stability issues is the inherent structural instability of halide perovskites. At room temperature, a common perovskite such as CsPbI\textsubscript{3} undergoes a structural phase transition from a perovskite phase to a more stable non-perovskite phase \cite{Protesescu2017,Misra2015}. Additionally, depending on the temperature and composition, perovskites can form several perovskite phases with varying optoelectronic properties \cite{Ma2019,Castelli2014}. In recent years, extensive experimental and theoretical efforts have been made to stabilize photoactive perovskite phases and increase the understanding of the mechanisms affecting their stability \cite{Ono2018,Wang2019,Asghar2017,Niu2015}. However, while experimental studies give valuable insights into the macroscopic properties of perovskite systems, they often lack the ability to provide a comprehensive understanding of their atomistic behaviour. Computational modelling has proven to be a valuable complement to experimental studies, allowing for atomic scale insights that are otherwise difficult to obtain \cite{Jiang2020,LaniganAtkins2021,Bouguima2021}.

Most computational studies of MHPs are performed using density functional theory (DFT) \cite{Zhang2021,Yaffe2017,Singh2020}. However, due the high computational costs of DFT it is unsuitable for the simulation of large systems over long timescales. Additionally, the considerable computation time restricts DFTs ability to easily investigate a range of MHP compositions at a multitude of temperatures. Computational methods based on classical force fields can provide a solution to the size and time constraints of DFT. Classical force fields have been employed successfully for the study of ion-mobility \cite{Barboni2018} and rotational dynamics \cite{Mattoni2015} of MAPbI\textsubscript{3}, as well as the dynamics of mixed CsPbI\textsubscript{1-x}Br\textsubscript{x} \cite{Balestra2020}. However, due to the classical nature of these force fields they are unable to describe electrons and ions simulateously, which are essential for the description of many MHP properties. Moreover, appropriate force fields have to be acquired for each material, which can be a complex task, especially when transferability is required.

Semi-empirical quantum mechanical methods, such as density functional tight-binding (DFTB), can help bridge the gap between DFT and classical force fields, and be an effective addition to these common computational methods \cite{Elstner2014,Loureno2020}. As DFTB is based on an approximation of the total DFT Kohn-Sham energy, it retains the ability to describe electrons and ions simultaneously, whilst significantly reducing computational cost. This is achieved by pre-calculating element pair interactions, which are then stored in Slater-Koster files \cite{Elstner1998}. However, creating the Slater-Koster files is a time consuming task, and they have limited transferability between elements. As a result, Slater-Koster files are unavailable for common perovskite elements including Cs, Pb, Sn, etc. GFN1-xTB, a recently developed extended tight-binding method, aims to solve this shortcoming by employing a reduced set of physically interpretative parameters, making parameterization a less complex task. As a result, \citet{Grimme2017}, the authors of this method, have published parameters for elements up to Z=86. A recent review of their performance for MHPs shows promising results \cite{VicentLuna2021}. However, a consistent overcompression of the perovskite systems remains.

In this work, we refine the GFN1-xTB parameters to accurately predict the perovskite phases of both CsPbBr\textsubscript{3} and CsPbI\textsubscript{3}, allowing for the analysis of both pure and mixed halide perovskites using a single set of parameters. To obtain a representative set of parameters, the GFN1-xTB parameters are trained against a set of DFT reference data. With the newly acquired parameters, we provide insights into MHP phase stability. To do so, we employ a combination of phase diagrams, heat maps, and mean squared displacements, and elaborate on the role of cation and halide dynamics on perovskite phase and stability. Additionally, we analyze the effect of mixed halide perovskite composition on structural properties and phase transition temperatures.

The new parameters were trained against a set of DFT reference data using the covariance matrix adaption evolution strategy (CMA-ES) optimization method \cite{Hansen2001} as implemented in the AMS2020 ParAMS module \cite{komissarov2021params,generalams}. The reference data was calculated with the BAND module of AMS2020 \cite{generalams,scmband} using the PBE-D3(BJ) exchange-correlation (XC) functional \cite{Perdew1997,Perdew1996,Johnson2005}. Training data included calculations on pure and mixed structures, i.e. relative energies upon minor distortions of the systems around their equilibrium configurations, octahedral tilting barriers, etc. (more details are provided in the Supporting Information). Starting from the original GFN1-xTB parameters \cite{Grimme2017} of Cs, Pb, Br, and I, we refined several parameters to obtain a better description of the MHPs. To avoid the general overcompression of the pervskite systems \cite{VicentLuna2021} we started optimizing the parameters of the repulsion energy term ($\alpha_{A}$ and $Z_{A}$), which is independent of the electrons. However, modifying this simple term is not sufficient for describing the properties of all the perovskite phases simultaneously. With halide perovskites being ionic crystals, electrostatic interactions are very important in these systems, thus, in addition to the repulsion term, we simultaneously modified the parameters accounting for the electrostatic energy contribution ($\eta_{A}$) of the electronic term. Finally we slightly modified the element pairs scaling parameters ($K_{A-B}$) of the zero-order electronic energy for a better description of all perovskite phases and chemical compositions.

\begin{figure}[!t]
    \centering
    \includegraphics[width=0.5\textwidth]{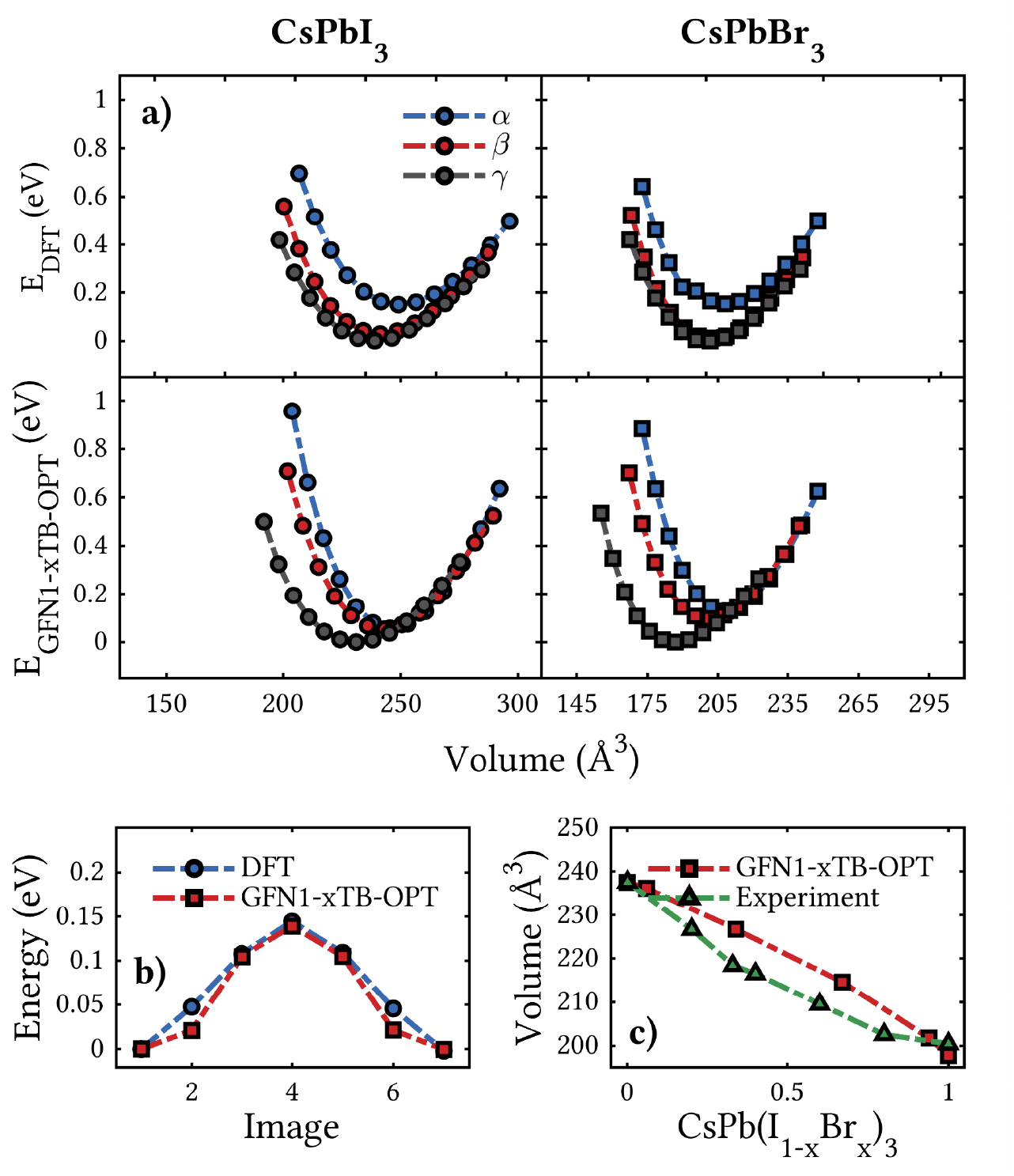}\\
    \caption{\textbf{a}: EOS of the perovskite phases of CsPbBr\textsubscript{3} (left) and CsPbI\textsubscript{3} (right), optimized with GFN1-xTB-OPT, and DFT. \textbf{b}: Octahedral tilting barrier of DFT optimized orthorhombic CsPbBr\textsubscript{3}, calculated with GFN1-xTB-OPT and DFT. \textbf{c}: Volume of CsPb(I\textsubscript{1-x}Br\textsubscript{x})\textsubscript{3} at 300K from GFN1-xTB-OPT molecular dynamics, and experiments \cite{Beal2016}.}
    \label{fig:structures}
\end{figure}

Following this procedure, a new set of parameters was obtained that can accurately reproduce reference and experimental data. From this point on, the new parameters will simply be referred to as GFN1-xTB-OPT for brevity, and their values can be found in Table S\ref{suptab:parameters}. To gain insight into the predictive performance of the GFN1-xTB-OPT parameters, we calculated properties they were not specifically trained against. A brief overview is given in Figure \ref{fig:structures}, where the results are compared to DFT calculations and experimental data. A more detailed examination, including GFN1-xTB-OPT performance on training set entries, and calculations with the original parameters, is available in the Supporting Information. The comparison between the GFN1-xTB-OPT and DFT equations of state (EOS) in Figure \ref{fig:structures}a shows good agreement between both the volumetric and energetic trends of the three perovskite phases. Differences in volume have been reduced from 8-38\% with the original parameters, to 0-7\% with GFN1-xTB-OPT. A closer look at the fully geometry optimized perovskite structures also shows minimal discrepancies between experiments and DFT, and GFN1-xTB-OPT optimized lattice parameters (Table S\ref{suptab:refvol}). These have been reduced to a difference of between 0 and 2.3\% for all perovskite phases, with the exception of orthorhombic CsPbBr\textsubscript{3}, where an overcompression of 8\% remains in one direction. Nevertheless, we have achieved a significant improvement over the underprediction of 37\% with the original parameters. Additionally, the new parameters show significant improvements with regard to the octahedral tilting barriers in the orthorhombic structures (Figure \ref{fig:structures}b), accurately predicting the in-phase and out-of-phase tilting in the orthorhombic phase. We found this to be critical for the reproduction of dynamical properties, since the octahedral tilting characterizes the differences between the orthorhombic, tetragonal, and cubic phases. Lastly, we performed molecular dynamics simulations of several mixed compositions, and found that our results predict the composition dependent volumetric expansion of CsPb(I\textsubscript{1-x}Br\textsubscript{x})\textsubscript{3} correctly, with a slight overprediction in the mixed perovskites.

\begin{figure}[!t]
    \centering
    \includegraphics[width=1.0\textwidth]{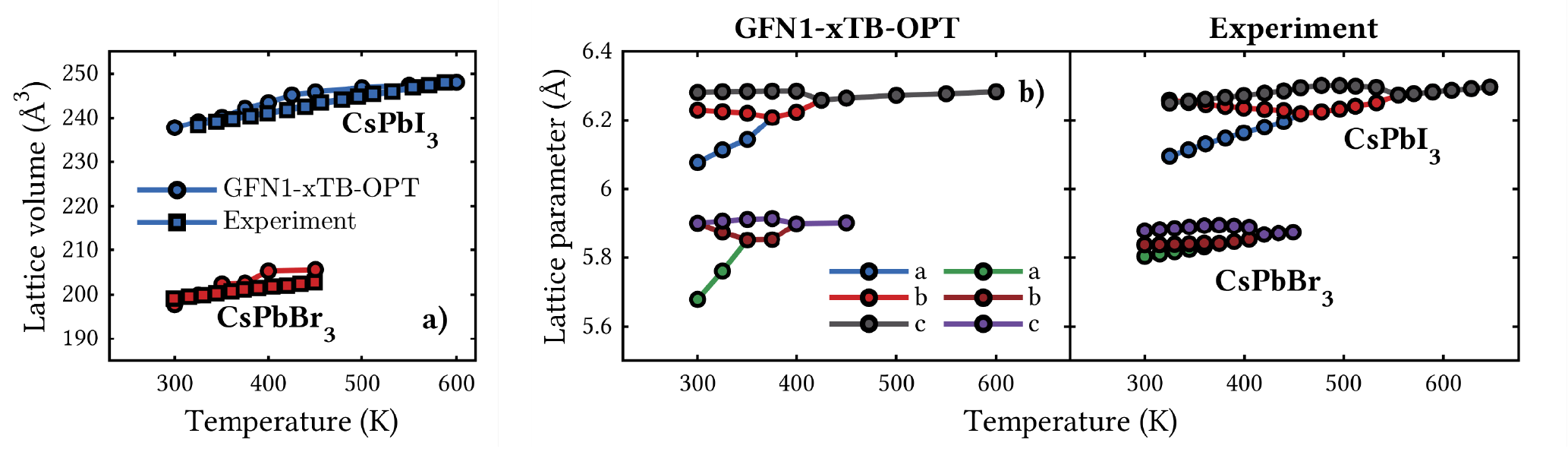}\\
    \caption{Volume (\textbf{a}) and phase diagram (\textbf{b}) of CsPbBr\textsubscript{3} and CsPbI\textsubscript{3} using GFN1-xTB-OPT and from experiments \cite{Stoumpos2013,Marronnier2018}.}
    \label{fig:phasedia}
\end{figure}

Using the newly acquired GFN1-xTB-OPT parameters, we first take a look at the volumetric expansion of pure CsPbBr\textsubscript{3} and CsPbI\textsubscript{3}. Molecular dynamics simulations were performed over a range of discrete temperatures between \SI{250}{\kelvin} and \SI{600}{\kelvin}, to obtain the temperature dependence of the lattice parameters and volume. Additional details on the simulation settings and lattice parameters extraction can be found in the Supporting Material). We found good agreement between the volumetric expansion of CsPbBr\textsubscript{3} and CsPbI\textsubscript{3} and experiments (Figure \ref{fig:phasedia}a), with the exception of a slight discrepancy for CsPbBr\textsubscript{3} in the orthorhombic phase. We attribute this to the overcompression of lattice parameter a in the orthorhombic phase, which we also found in previously performed geometry optimizations. 

Next, we look more specifically at the phase behaviour of the perovskites in Figure \ref{fig:phasedia}b. From the evolution of the lattice parameters we found that both CsPbBr\textsubscript{3} and CsPbI\textsubscript{3} are in the orthorhombic phase at \SI{300}{\kelvin}. For increasing temperatures, \SI{350(13)}{\kelvin} and \SI{400(13)}{\kelvin}, CsPbBr\textsubscript{3} transitions to the tetragonal and cubic phases respectively. Similarly, two phase transitions occur at \SI{400(13)}{\kelvin} and \SI{425(13)}{\kelvin} in CsPbI\textsubscript{3}. This phase behaviour is in qualitative agreement with experiments \cite{Stoumpos2013,Marronnier2018}. However, the transition temperatures are shifted, most notably in CsPbI\textsubscript{3}. Additionally, the tetragonal phase in CsPbI\textsubscript{3} is found over a shorter range of temperatures. We attribute this to fluctuations between the cubic and tetragonal phase during the simulation, which occur occasionally. Additionally, in the tetragonal phase, the tilting direction often varies during the simulation, resulting in on average cubic lattice parameters. We think the behaviour can be improved by using larger cells, as we have observed similar but more pronounced behavior in smaller systems. There, the tetragonal phase was completely absent in the phase diagram of CsPbI\textsubscript{3} as a result of fluctuations in tilting direction (Figure S\ref{supfig:smallersystem}). Nevertheless, using a single set of parameters, we correctly reproduce the general phase behavior of both materials, as well as the qualitative difference in phase transition temperatures between them.

\begin{figure}[!t]
    \centering
    \includegraphics[width=\textwidth]{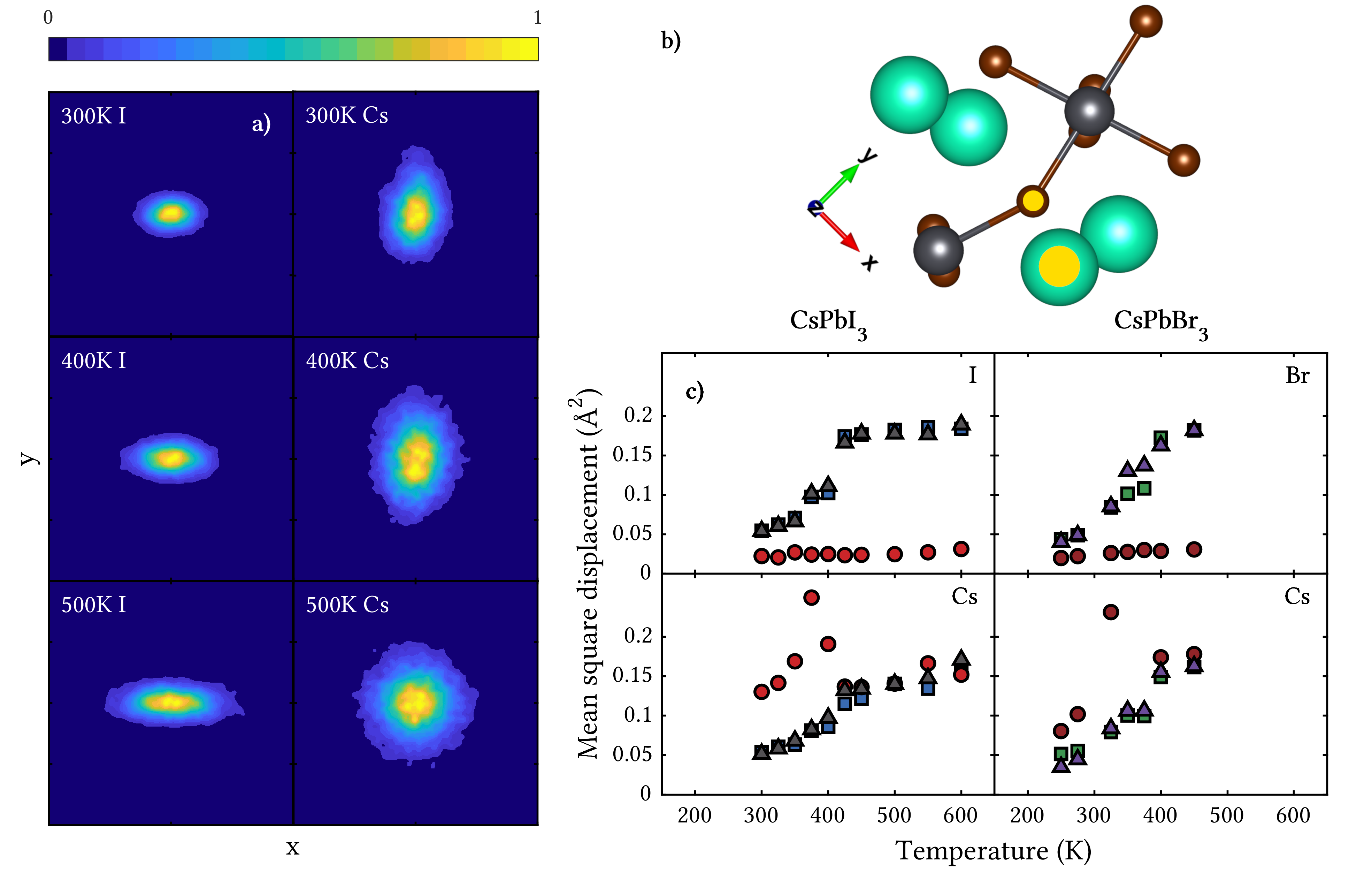}\\
    \caption{\textbf{a}: Heat maps of I (left) and Cs (right) in CsPbI\textsubscript{3} at several temperatures, each corresponding to one perovskite phase the system is in. \textbf{b}: Schematic overview of the halide and cation that were used for the heat maps and MSD calculations. \textbf{c}: MSD of X (X = Br, I) (top), and Cs (bottom) in CsPbBr\textsubscript{3} and CsPbI\textsubscript{3} along three directions over a range of temperatures.}
    \label{fig:heatmsd}
\end{figure}

Having analyzed the general phase behavior of CsPbBr\textsubscript{3} and CsPbI\textsubscript{3}, we now look in more detail at the displacement of individual ions, and its impact on perovskite phase and phase stability. To do so, we studied positional heat maps of I and Cs ions in CsPbI\textsubscript{3} (Figure \ref{fig:heatmsd}a), as well as the MSD in all directions of X (X = Br, I) and Cs in CsPbBr\textsubscript{3} and CsPbI\textsubscript{3} (Figure \ref{fig:heatmsd}c). A schematic overview is shown in Figure \ref{fig:heatmsd}b. 

We first focus on the heat maps of I at \SIlist{300;400;500}{\kelvin}. At \SI{300}{\kelvin}, only a slight displacement of the halides is observed, with a preferential direction towards the surrounding cations. This behaviour can be readily explained by the relation between phase stability and cation-halide contact \cite{Goldschmidt1927,Bechtel2018}. In the orthorhombic phase, the tilting of the octahedra, together with the displacement of the Cs cation toward the edge of the octahedral cage, increases the cation-halide contact, which stabilizes the perovskite. The minimal displacement of the halides results in optimal contact. As the thermal energy increases and the system transitions to the tetragonal and cubic phases (\SIlist{400;500}{\kelvin}), the halides fluctuate more rapidly between the surrounding cations, which, in these phases, have moved towards the center of the octahedral cage. This is quantified in the plots of the MSDs, where significant jumps are found in MSD between the different phases, with overall anisotropic halide displacement. A similar trend is found in CsPbBr\textsubscript{3}, with the MSD increasing at each phase transition. We suggest that these fluctuations play a significant role in stabilizing the tetragonal and cubic perovskite phase. A recent experimental report by \citet{Songvilay2019} supports these findings, showing a strong increase in anisotropy of the halide movement for rising temperatures.

Next, we analyze the heat maps of the Cs cation at the same temperatures (Figure \ref{fig:heatmsd}a). At \SI{300}{\kelvin}, we observe preferential movement of the Cs cation towards the positive y-direction. We suggest that the tendency of the cations to move away from their equilibrium position near the edge of the cage, towards the center of the cage, decreases the cation-halide contact, leading to two possible effects. First, at low temperatures, in the orthorhombic phase, we propose that the decreased cation-halide contact can induce a transition to the non-perovskite yellow phase. These findings corroborate observations by \citet{Straus2020}, who found similar preferential movements using single-crystal X-ray diffraction measurements. Second, we find significant increases in the anisotropic cation displacement upon the orthorhombic to tetragonal phase transition, where the equilibrium position of the Cs cation is moved towards the center of the cage. This actually improves the cation-halide contact, as the cations fluctuate predominantly between the halides that are tilted away from their equilibrium positions. We suggest that, once enough thermal energy is available to the system, the loss of contact from the directional Cs displacement in the orthorhombic phase can be compensated for by increased halide and cation fluctuations, therefore initiating the transition to the tetragonal phase. At \SI{500}{\kelvin}, the Cs fluctuations have become completely isotropic around the equilibrium positions. While this results in a cubic structure on average, the system exhibits many local distortions at any given moment. We propose that the substantial halide and cation fluctuations in the tetragonal and cubic phase help stabilize these perovskite phases. This is in contrast to the destabilizing effect of the Cs displacement in the orthorhombic phase.

\begin{figure}[!t]
    \centering
    \includegraphics[width=\textwidth]{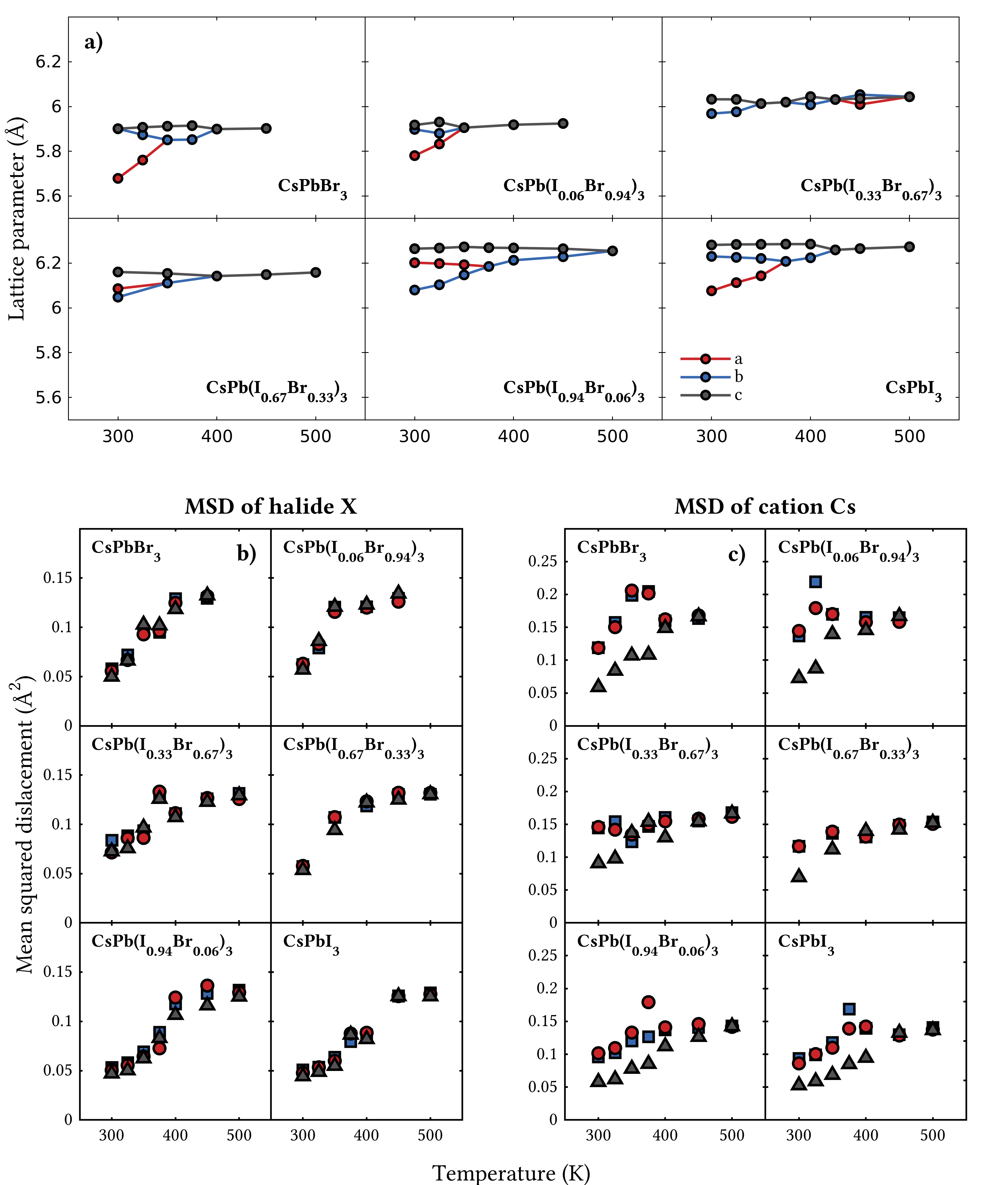}\\
    \caption{\textbf{a}: Phase diagrams of several CsPb(I\textsubscript{1-x}Br\textsubscript{x})\textsubscript{3} mixed and pure compositions, simulated with GFN1-xTB-OPT. \textbf{b} and \textbf{c}: MSD of all halides (\textbf{b}) and Cs cations (\textbf{c}) from the same mixed and pure perovksite compositions.}
    \label{fig:phasediamix}
\end{figure}

With previous observations in mind, we analyze the effects of mixing halides on the phase diagrams. Simulations are done for four CsPb(I\textsubscript{1-x}Br\textsubscript{x})\textsubscript{3} compositions: x = 0.06, 0.33, 0.67, 0.94. The phase diagrams are plotted in Figure \ref{fig:phasediamix}a, along with those for the pure systems for comparison. Starting with the more almost pure compositions, x = 0.94 (top middle) and x = 0.06 (bottom middle), we observe no significant change in phase transition temperatures, but a decrease in the orthorhombic nature of the structures at \SIlist{325;350}{\kelvin} respectively, which is evident from the decrease in a/b ratio from xx to xx. A change in a/b ratio indicates a change in octahedral tilting, and it converges to 1 in the tetragonal phase. Although the lattice parameters indicate a direct orthorhombic to cubic phase transition for this composition, octahedral tilting is still clearly present in the averaged structures, as is shown in the supporting material (Figure \ref{supfig:tetragonal}). In the more mixed compositions, x = 0.33 and x = 0.67 (top and middle right), we observe a significant change in the phase diagrams, with the orthorhombic to tetragonal phase transition shifted to lower temperatures: \SIlist{350;300}{\kelvin} respectively. However, a closer look at the averaged structures of x = 0.67 at \SI{300}{\kelvin} again reveals that the local structure does not match with the expectation from lattice parameter observations, revealing a more orthorhombic like structure. Generally, ionic movement throughout the simulation is more disordered in these mixed compositions, resulting in locally distorted structures when averaged over a longer time. As a result, the global fluctuations are in some cases more difficult to directly relate to the local structure of the perovskite. Nevertheless, a clear decrease in orthorhombic nature of the structure can be seen for all mixed halide perovskites. 

We suggest these findings to be the result of increased halide fluctuations present in mixed perovskites, shown in Figure \ref{fig:phasediamix}b. Here, we calculated the average MSD of all halide and Cs ions in the system. A clear increase in MSD is found at lower temperatures for x = 0.67 (\SIrange{300}{375}{\kelvin}) and x = 0.33 (\SIrange{325}{400}{\kelvin}) compared to the pure perovskites closest in composition. For x = 0.94 at \SI{350}{\kelvin} and x = 0.06 at \SI{400}{\kelvin}, similar increases are found. Additionally, at these temperatures, the MSD of the Cs cations increases as well, often with more isotropic displacement (e.g. x = 0.33 at \SIrange{350}{400}{\kelvin} in Figure \ref{fig:phasediamix}c). We suggest that these fluctuations, together with the decreased volume of the mixed structures compared to pure CsPbI\textsubscript{3}, result in better contact between the cations and halides, stabilizing the mixed structures at decreased temperatures. We highlight that such stabilizing effect can be achieved even with a slight modification of the compositions of the perovskites, for example in the case of x = 0.94 at \SI{350}{\kelvin}, and x = 0.06 at \SI{400}{\kelvin}.  

In conclusion, we have re-parameterized a set of GFN1-xTB parameters to accurately describe the structures and dynamics of CsPbBr\textsubscript{3}, CsPbI\textsubscript{3}, and CsPb(I\textsubscript{1-x}Br\textsubscript{x})\textsubscript{3}. Using this method, we studied the effect of ion dynamics on the phase stability of pure and mixed Cs-based halide perovskites. At low temperatures, in the orthorhombic phase, the Cs cation is located at the edge of the octahedral cage, providing good contact with the surrounding halides. However, we show that Cs tends to move away from this position, decreasing the cation-halide contact, and potentially resulting in degradation to the non-perovskite yellow phase. At higher temperatures, the decrease in cation-halide contact can be compensated by increased halide fluctuations, which initiates the transition to the tetragonal phase. Interestingly, our simulations of several mixed halide perovskites demonstrate a consistently less orthorhombic nature and/or lower phase transition temperatures compared to pure systems. This can be explained by the fact that at similar temperatures, halide fluctuations are more pronounced in the mixed perovskites than in the pure systems, and Cs cations movement is more isotropic. Both effects increase the cation-halide contact and therefore improve perovskite phase stability. We highlight that such stabilization effect is already significant when mixing small amounts of halides, such as 6\% of Br in CsPbI\textsubscript{3} or I in CsPbBr\textsubscript{3}. Our results provide a basis for the rational design of phase-stable halide perovskites.

\begin{acknowledgement}

S.R. and S.T. acknowledge funding by the Computational Sciences for Energy Research (CSER) tenure track program of Shell and NWO (Project No. 15CST04-2); J.M.V.L. and S.T. acknowledge NWO START-UP from the Netherlands. The authors also thank stimulating discussions within the SCALEUP consortium (SOLAR-ERA.NET Cofund 2, id: 32).

\end{acknowledgement}

\begin{suppinfo}

\section{1. Computational settings}
\subsection{1.1 DFT calculations}
All reference DFT calculations calculations, with the exception of the octahedral tilting barriers, were performed using the BAND module of the AMS software package \cite{scmband,generalams}. The PBE \cite{Perdew1996,Perdew1997} exchange-correlation (XC) functional was used, with the addition of the D3(BJ) \cite{Grimme2011} dispersion correction to account for long-range van der Waals interactions. The triple zeta + double polarization (TZ2P) basis sets were used, with core orbitals up to Cs-4d, Pb-4d, I-4p, and Br-3p considered as frozen during the self consistent field (SCF) procedure. 

For the EOS of pure and mixed structures in the training set, full geometry optimizations were first performed on all pure perovskite phases to obtain the fully relaxed structures. The numerical quality was set to basic, as it had minimal effects on final geometries and energies, while significantly increasing computational costs. The amount of k-points were set according to the size of each lattice vector: 9x9x9 for CsI and CsBr, 9x11x7 for PbI\textsubscript{2}, 9x7x7 for PbBr\textsubscript{2}, 9x9x9 for cubic ($\alpha$) CsPbBr\textsubscript{3} and CsPbI\textsubscript{3}, 7x7x9 for tetragonal ($\beta$) CsPbBr\textsubscript{3} and CsPbI\textsubscript{3}. Due to the larger system size, as well as the computational increase associated with an increased number of k-points, the geometry optimization of orthorhombic ($\gamma$) CsPbBr\textsubscript{3} and CsPbI\textsubscript{3} was performed with 5x3x5 k-points. Single point calculations for the energies and forces included in the training set were then performed with the k-points increased to 7x5x7. During the geometry optimizations, the ionic positions, as well as the lattice vectors were allowed to change. The optimization was performed using the fast inertial relaxation method (FIRE) \cite{Bitzek2006}, with the energy, gradients, and stress energy per atom convergence criteria set to \SI{1e-5}{\hartree\per\angstrom}, \SI{1e-3}{\hartree\per\angstrom}, and \SI{5e-4}{\hartree} respectively.

Pure, GFN1-xTB optimized perovskite structures were part of the training set as well. These calculations were performed with the original GFN1-xTB parameters, as implemented in the DFTB module of AMS \cite{dftbcite,generalams}. Identical convergence settings were used for the FIRE optimizer. After a full relaxation, single point DFT calculations were performed on several structural images taken from the optimization, with settings identical to those in used in previous DFT calculations.

NEB calculations on orthorhombic structures were done using the VASP simulation package \cite{Perdew2005,Kresse1993,Kresse19961,Kresse19962}. Calculations were performed using the same PBE XC functional with a D3(BJ) dispersion correction. The plane-wave basis set was used to an energy cutoff of 500 eV, and the energy convergence was set to \SI{0.1}{\milli\electronvolt}. Five intermediate images were used, which were allowed to change in volume during the NEB image calculation. To keep the energies of this subset of the training data consistent with the aforementioned data, single point calculations were performed with the BAND engine of AMS on the final geometries obtained form the NEB calculations, using previously defined settings.

\subsection{1.2 Molecular dynamics simulations}
Molecular dynamics simulations were all done in the AMS2020 software package \cite{generalams}. Simulations were performed at static temperatures, over a large temperature range. Equilibration runs were done at each temperature, starting from a GFN1-xTB-OPT optimized orthorhombic 2x2x2 supercell with 160 ions. The equilibrations ran for \SI{100}{\pico\second} in an NPT-ensemble, using a Berendsen thermostat and barostat with a time-step of \SI{0.5}{\femto\second}. Afterwards, production runs were done for another \SI{100}{\pico\second} in an NPT-ensemble, using the Nosé-Hoover (NH) \cite{Hoover1985,Nose1984} thermostat and Martyna-Tuckerman-Tobias-Klein (MTTK) \cite{Martyna1994,Martyna1996} barostat, with a time-step of \SI{1}{\femto\second}.

To reduce the fluctuations present in the systems, and allow for increased sampling of ions for the mean squared displacement (MSD) and heat maps, the system sizes for the runs in this work were then increased to a 2x3x3 orthorhombic supercell with 360 ions. The time-step was increased to \SI{2}{\femto\second}, which had no significant effects on system energy or behaviour. This was tested by performing MD simulations of identical systems using different time-step values, from which we obtained the same energy and lattice parameters fluctuations. Simulations were then performed for another \SI{100}{\pico\second} using the NH and MTTK thermostat and barostat, starting from the previously equilibrated systems. For both the equilibration and production runs, the damping constants were set to $\tau_t=\SI{100}{\femto\second}$ and $\tau_p=\SI{1500}{\femto\second}$ for the thermostat and barostat respectively. A charge convergence of \SI{1e-5}{\hartree} in the self-consistent cycle, a DIIS of 10, and a Slater radial threshold of \SI{1e-3}{} were used as settings for the DFTB engine in AMS \cite{generalams,dftbcite,DFTB2020}.

Lattice parameters were extracted by means of a Gaussian fit. Data was sampled from the simulation each \SI{24}{\femto\second}, and the results were calculated by fitting the data from the last \SI{50}{\pico\second} of each simulation.

\section{2. Mean squared displacements}
Due to the asymmetric nature of the orthorhombic perovskite structure, which contains four and twelve unique sites for Cs and I or Br respectively, only the 16 equivalent ions of each species present in the simulated systems were used for the MSD calculations and heat maps. Mean squared displacements were then determined from the positions of these ions at each step. To do so, the most occupied position was first determined for each ion with a Gaussian fit, after which the mean squared displacement from this position is determined along each direction using:

\begin{equation}
    MSD=\frac{1}{t}\sum_{i=0}^t\left|x^{(i)}-x_A\right|^2,
\end{equation}

\noindent where $t$ is the number of data points saved during the simulation, $x^{(i)}$ the position at time $i$, and $x_A$ the most occupied position. The results are then averaged over all equivalent ions. The ions in a pristine perovskite crystal without any crystal defects only fluctuate around their equilibrium positions. As such, they ions do not diffuse throughout the material, and the calculated MSD is a measure of the magnitude of ion fluctuations, not migration throughout the bulk.

\section{3. Lattice parameter fitting}
Due to the significant fluctuations in lattice parameters during the constant temperature MD simulation, it is hard to uniquely determine the magnitude of the lattice parameters. Additionally, close to phase transition temperatures, the perovskites can occasionally fluctuate between two phases. We therefore closely follow a method outlined by \citet{Jinnouchi2019} to determine the lattice parameters for each run.

First, the pseudo-cubic lattice parameters are binned, after which a Gaussian function is fitted over the resulting distribution. The Gaussian distribution is defined as follows:

\begin{equation}
    \rho\left(x\right)=\frac{1}{\sqrt{2\pi\sigma^2}}e^{\left(-\frac{\left(x-\mu\right)}{2\sigma^2}\right)},
\end{equation}

\noindent where $x$ is the binned lattice parameters, and $\mu$ and $\sigma$ are the equilibrium pseudo-cubic lattice parameter and their variance respectively. Initially, all three lattice parameters are determined individually by fitting the Gaussian distribution. If the difference between the equilibrium lattice parameters is larger than $\sqrt{\left(\sigma_1^2+\sigma_2^2+\sigma_3^2\right)/3}$, where each $\sigma_i \left(i=1,2,3\right)$ corresponds to a lattice parameter, the structure is determined to be orthorhombic. However, if it is smaller, the two lattice parameters with a smaller difference are grouped. Two Gaussians are then fitted over the new distributions. If the difference between the results is larger than $\sqrt{\left(\sigma_1^2+\sigma_2^2\right)/2}$, the structure is deemed tetragonal. If not, all lattice parameters are grouped again, and a final fit is performed.

\section{4. Training set}
As mentioned before, the training set contained several types of reference data. Included were all pure CsPbBr\textsubscript{3} and CsPbI\textsubscript{3} perovskite phases, isotropically stretched and compressed 10\%. The resulting EOS, as well as single point calculations with the original GFN1-xTB, and GFN1-xTB-opt parameters, are shown in Figure \ref{supfig:eqosdft}. Similarly stretched and compressed precursor structures (CsI, CsBr, PbI\textsubscript{2}, and PbBr\textsubscript{2}) were included as well. For the perovskite structures, the formation energies and interatomic forces were used in the training set; and for the precursor structures we added the energies relative to the equilibrium structure, as well as the interatomic forces.

\begin{figure}
    \centering
    \includegraphics[width=0.65\textwidth]{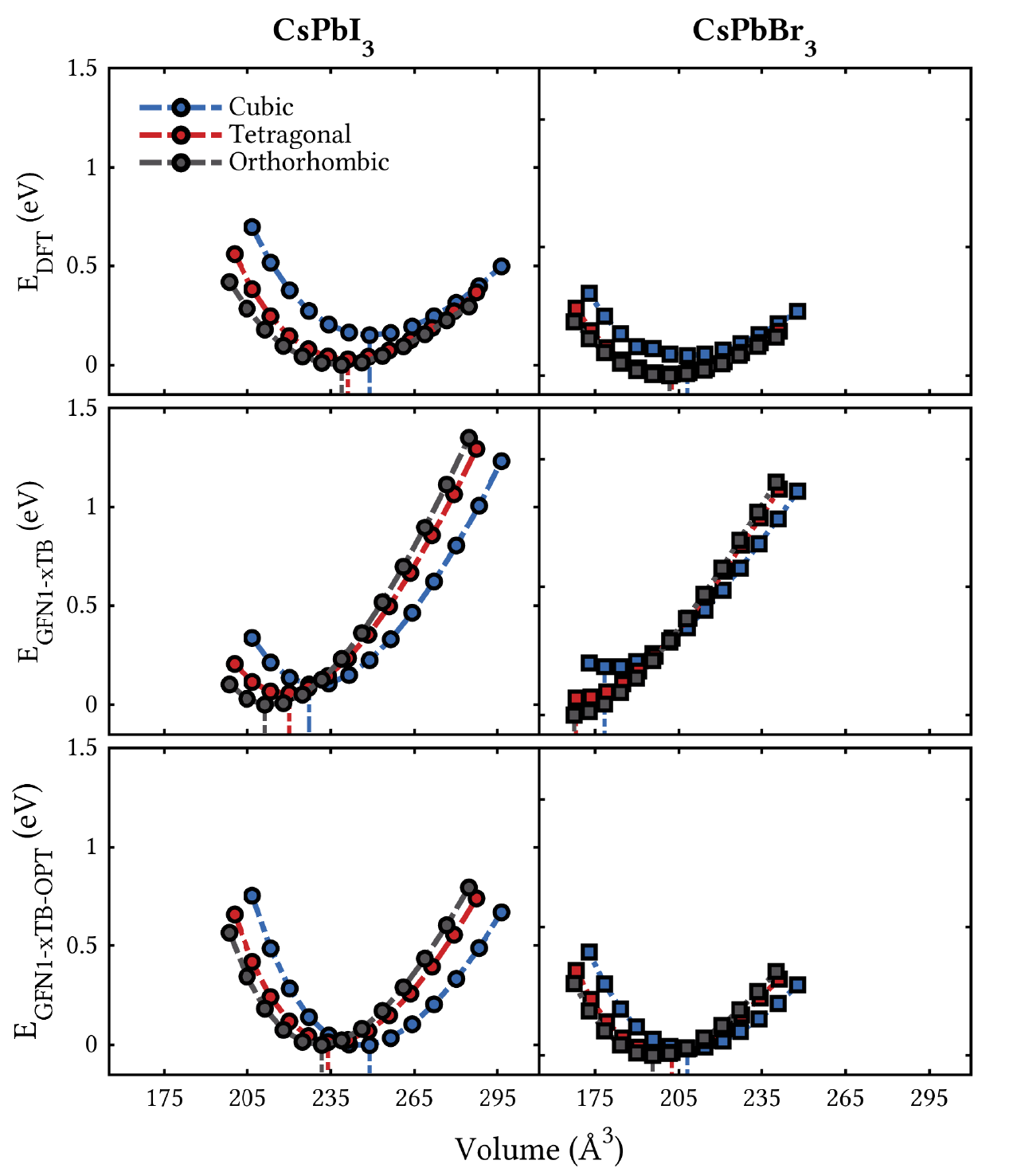}\\
    \caption{EOS of the perovskite phases of CsPbBr\textsubscript{3} (left) and CsPbI\textsubscript{3} (right), calculated with single point GFN1-xTB, GFN1-xTB-OPT, and DFT calculations on DFT optimized structures. Full geometry optimizations were performed with DFT to obtain the equilibrium structure, after which they were isotropically stretched and compressed.}
    \label{supfig:eqosdft}
\end{figure}

Additionally, octahedral tilting barriers were used in the parameter optimization. Both in-phase and out-of-phase tilting barriers were added for CsPbI\textsubscript{3} and CsPbBr\textsubscript{3}. Used as training data were the energies relative to the orthorhombic equilibrium structures, and the interatomic forces. Both the reference, and single point GFN1-xTB-OPT calculations are given in Figure \ref{fig:NEBdft}.

\begin{figure}
    \centering
    \includegraphics[width=\textwidth]{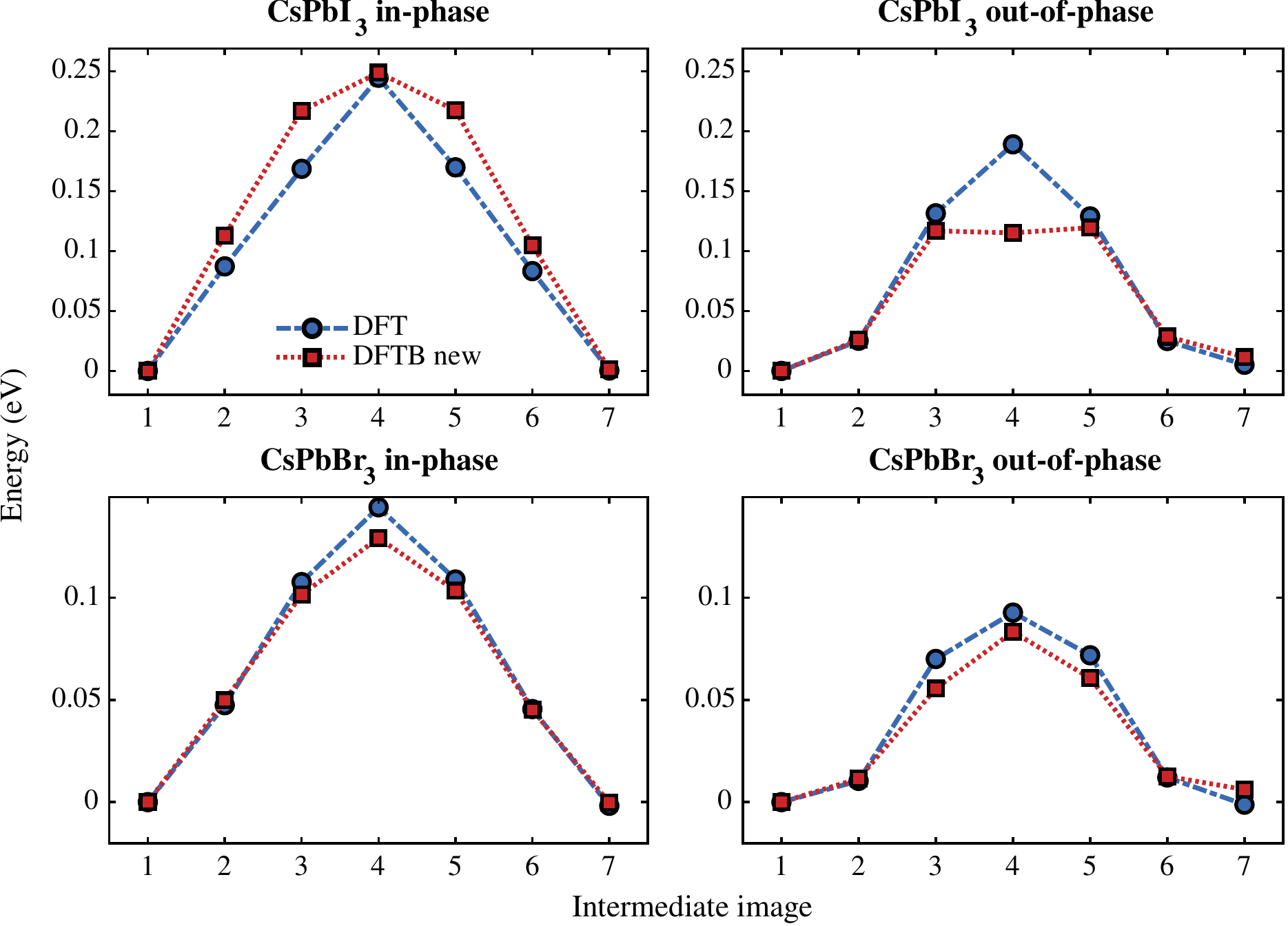}\\
    \caption{In-phase and out-of-phase octahedral tilting barriers of CsPbBr\textsubscript{3} (top) and CsPbI\textsubscript{3} (bottom). Structural images were first calculated with VASP in a full NEB calculation, after which single point calculations were performed with GFN1-xTB-OPT and DFT.}
    \label{fig:NEBdft}
\end{figure}

Lastly, several cubic mixed halide perovskite compositions were used, as well as structural images from GFN1-xTB optimizations of the perovskite phases. The mixed halide perovskites were stretched and compressed isotropically only 2.5\%, adding their relative energies and interatomic forces. For the GFN1-xTB optimized structures, the formation energies and interatomic forces were used. 

\newpage

\section{Supporting Tables}

\def\arraystretch{1.5}
\begin{table}
\caption{Optimized GFN1-xTB-OPT parameters for CsPbI\textsubscript{3} and CsPbBr\textsubscript{3}.}
\begin{tabular}{|c|c|}
\hline
Parameter & Value \\ \hline
$\alpha_{Cs}$   & 0.661 \\ \hline
$\alpha_{Pb}$   & 1.043 \\ \hline
$\alpha_{Br}$   & 0.490 \\ \hline
$\alpha_{I}$    & 0.673 \\ \hline
$Z_{Cs}$        & 19.945 \\ \hline
$Z_{Pb}$        & 19.89 \\ \hline
$Z_{Br}$        & 8.816 \\ \hline
$Z_{I}$         & 78.403 \\ \hline
$\eta_{Br}$     & 0.367 \\ \hline
$\eta_{I}$      & 0.298 \\ \hline
$K_{Cs-Cs}$     & 1.020 \\ \hline
$K_{Cs-Pb}$     & 0.988 \\ \hline
$K_{Cs-Br}$     & 1.017 \\ \hline
$K_{Cs-I}$      & 0.980 \\ \hline
$K_{Pb-Pb}$     & 0.983 \\ \hline
$K_{Pb-Br}$     & 1.018 \\ \hline
$K_{Pb-I}$      & 0.993 \\ \hline
$K_{Br-Br}$     & 0.989 \\ \hline
$K_{Br-I}$      & 1.015 \\ \hline
$K_{I-I}$       & 1.020 \\ \hline
\end{tabular}
\label{suptab:parameters}
\end{table}

\def\arraystretch{1.5}
\begin{table}
\caption{Lattice vectors and volume of the CsPbX\textsubscript{3} perovskite phases obtained from full geometry optimization with GFN1-xTB, GFN1-xTB-OPT, and DFT; and from experiments.}
\begin{tabular}{| c | c |l|c|c|c|c|}
\hline
Material                 & Phase                         &  & GFN1-xTB & GFN1-xTB & DFT & Experiment \\[-1ex] & & & & -OPT & & \\ \hline
CsPbI\textsubscript{3}             & Cubic                         & a, b, c (\r{A})     & 6.12     & 6.26           & 6.29  & 6.18 - 6.30      \\
                         &                               & V (\r{A}$^3$)       & 230      & 245            & 249   & 236 - 250 \cite{Marronnier2018,Beal2016,Trots2008,Eperon2015}        \\ \cline{2-7}
                         & Tetragonal                    & a, b (\r{A})        & 8.44     & 8.78           & 8.69  & 8.83              \\
                         &                               & c (\r{A})           & 6.19     & 6.30           & 6.39  & 6.30              \\
                         &                               & V (\r{A}$^3$)       & 441      & 486            & 482   & 491 \cite{Marronnier2018}              \\ \cline{2-7}
                         & Orthorhombic                  & a (\r{A})           & 9.05     & 8.81           & 8.93  & 8.82 - 8.86       \\
                         &                               & b (\r{A})           & 7.03     & 8.38           & 8.59  & 8.58 - 8.65       \\
                         &                               & c (\r{A})           & 11.78    & 12.52          & 12.46 & 12.47 - 12.52     \\
                         &                               & V (\r{A}$^3$)       & 749      & 924            & 956   & 947 - 955 \cite{Marronnier2018,Sutton2018}        \\ \hline
CsPbBr\textsubscript{3}            & Cubic                         & a, b, c (\r{A})     & 5.65     & 5.93           & 5.93  & 5.84 - 5.87       \\
                         &                               & V (\r{A}$^3$)       & 180      & 208            & 208   & 200 - 203 \cite{Stoumpos2013,Cottingham2016,Moller1959}       \\\cline{2-7}
                         & Tetragonal                    & a, b (\r{A})        & 7.51     & 8.22           & 8.22  & 8.25 - 8.29       \\
                         &                               & c (\r{A})           & 5.66     & 5.97           & 6.00  & 5.88 - 5.90       \\
                         &                               & V (\r{A}$^3$)       & 320      & 403            & 405   & 402 - 404 \cite{Stoumpos2013,Moller1959}        \\ \cline{2-7}
                         & Orthorhombic                  & a (\r{A})           & 8.69     & 8.47           & 8.33  & 8.26 - 8.32       \\
                         &                               & b (\r{A})           & 5.14     & 7.56           & 8.21  & 8.21 - 8.26       \\
                         &                               & c (\r{A})           & 11.09    & 11.68          & 11.79 & 11.60 - 11.77     \\
                         &                               & V (\r{A}$^3$)       & 495      & 748            & 807   & 793 - 798 \cite{Stoumpos2013,Atourki2017,Cottingham2016}        \\ \hline
\end{tabular}
\label{suptab:refvol}
\end{table}

\newpage

\section{Supporting figures}
\begin{figure}
    \centering
    \includegraphics[width=0.65\textwidth]{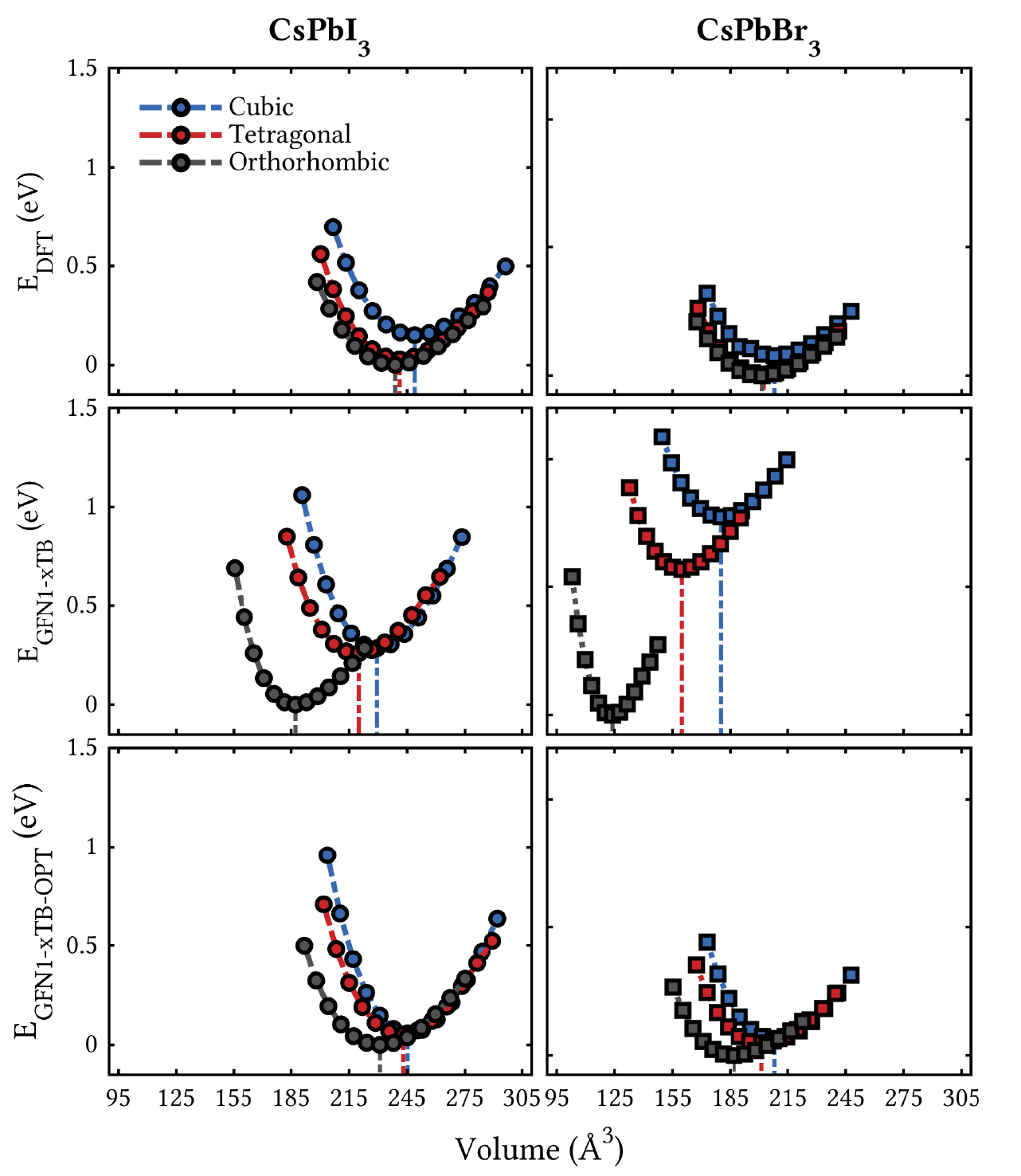}\\
    \caption{EOS of the perovskite phases of CsPbBr\textsubscript{3} (left) and CsPbI\textsubscript{3} (right), calculated with GFN1-xTB, GFN1-xTB-OPT, and DFT. Full geometry optimizations were performed with each method to obtain the equilibrium structure, after which they were isotropically stretched and compressed.}
    \label{supfig:eqosfull}
\end{figure}

\begin{figure}
    \centering
    \includegraphics[width=0.5\textwidth]{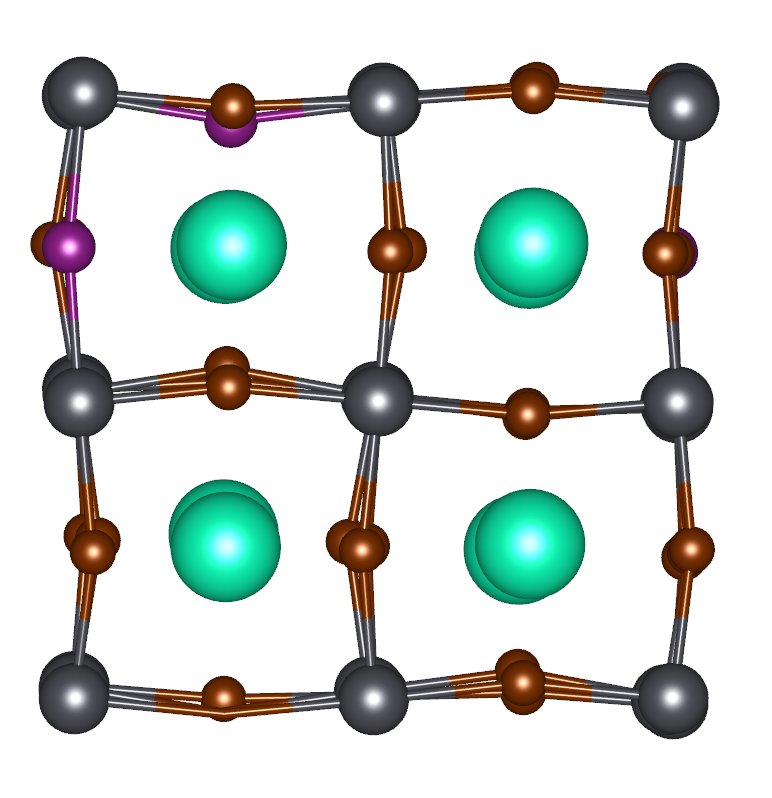}
    \caption{Averaged CsPb(I\textsubscript{0.06}Br\textsubscript{0.94})\textsubscript{3} structure from a \SI{350}{\kelvin} MD simulation.}
    \label{supfig:tetragonal}
\end{figure}

\begin{figure}
    \centering
    \includegraphics[width=\textwidth]{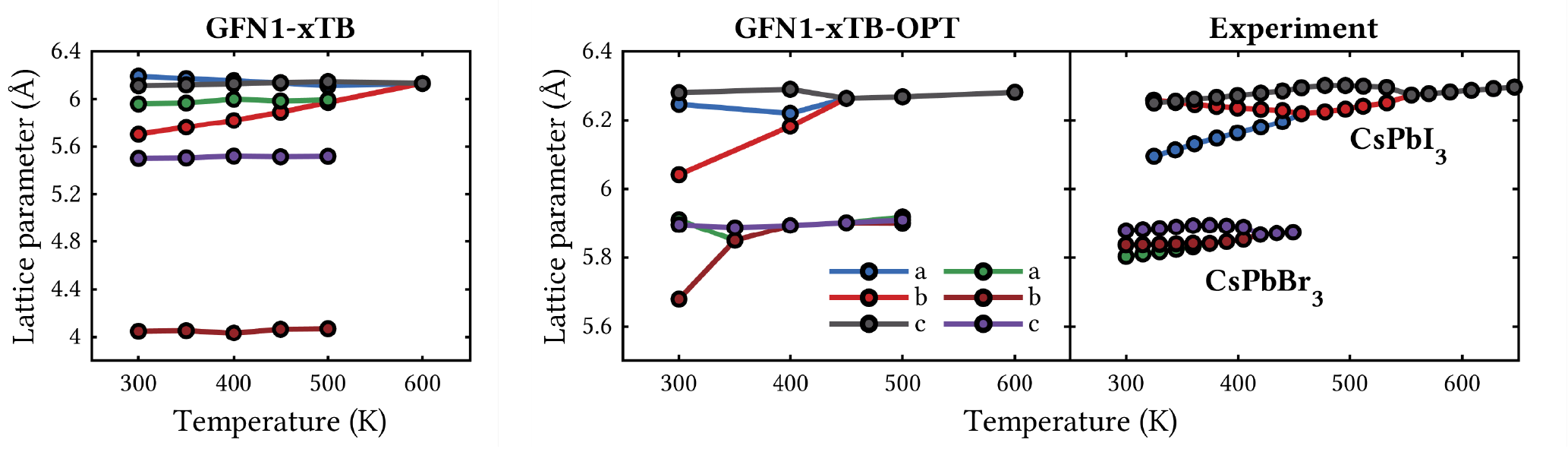}
    \caption{Phase diagram of CsPbBr\textsubscript{3} and CsPbI\textsubscript{3} from 180 ion large MD simulations with GFN1-xTB (left) and GFN1-xTB-OPT (middle), and from experiments (right) \cite{Stoumpos2013,Marronnier2018}.}
    \label{supfig:smallersystem}
\end{figure}

\end{suppinfo}

\newpage

\bibliography{ms}

\end{document}